\documentclass[reprint, amsmath, amssymb, aps]{revtex4-2}

\usepackage{graphicx}
\usepackage[utf8]{inputenc}
\usepackage[T1]{fontenc}
\usepackage{cmap}
\usepackage{color}
\usepackage{float}
\DeclareMathAlphabet{\pazocal}{OMS}{zplm}{m}{n}

\def\JCAPstyle#1{}
\def\Abstract#1
\def\imo{i}

\usepackage{soul}

\begin{document}

\title{Circumventing Quantum Gravity: Black Holes Evaporating into Macroscopic Wormholes}

\author{S. V. Bolokhov}\email{bolokhov-sv@rudn.ru}
\affiliation{Peoples' Friendship University of Russia (RUDN University),\\ 6 Miklukho-Maklaya Street, Moscow, 117198, Russia}

\author{R. A. Konoplya}\email{roman.konoplya@gmail.com}
%\affiliation{\blue{RUDN University, 6 Miklukho-Maklaya Street, Moscow, 117198, Russia}}
\affiliation{Research Centre for Theoretical Physics and Astrophysics, Institute of Physics, Silesian University in Opava, Bezru\v{c}ovo n\'am\v{e}st\'i 13, CZ-74601 Opava, Czech Republic}

\begin{abstract}
Recently, metrics that describe regular black holes, extreme black holes, or traversable wormholes have been widely discussed. These spacetimes, appearing in scenarios such as the brane world, are contingent on the values of the parameters, with each metric encompassing all three objects. We are considering various known models for these black hole/wormhole interpolating spacetimes and showing that, starting from the macroscopic black holes, all of them must evaporate into macroscopic wormholes, thus avoiding existential problems related to the final stages of black hole evaporation and issues of quantum gravity and black hole remnants. For this purpose, we are calculating the energy emission rates of black holes and the appropriate lifetimes. We argue that some of our conclusions should hold regardless of the specific model, as long as it permits an extremal black hole state with zero temperature at a particular value of the coupling constant.
\end{abstract}

\maketitle

\section{Introduction}\label{Intro}

Since Stephen Hawking's discovery of radiation emitted from the vicinity of a black hole's event horizon \cite{Hawking:1975vcx}, a vast number of studies have focused on the fate of black holes as they evaporate, particularly the final stages of this process, which may involve miniature black holes requiring spacetime quantization.

An initially rotating black hole radiates away its angular momentum through superradiance, eventually leading to a spherically symmetric configuration in the later stages of evaporation. As a Schwarzschild black hole loses mass through Hawking radiation—where quantum effects cause particle-antiparticle pairs near the event horizon to result in one particle escaping—the black hole's temperature increases as its mass decreases. This inverse relationship means that smaller black holes become hotter and emit radiation more rapidly, causing the evaporation process to accelerate as the black hole's mass diminishes.

However, this is not necessarily the case for Reissner-Nordstr\"om-like solutions, where the "charge" (which could also be interpreted differently, such as a tidal force from extra dimensions) and mass may radiate at different rates. If, for some reason, the charge or effective "charge" is lost more slowly than the mass, their ratio would increase, potentially reaching an extreme value where the black hole’s temperature vanishes, and evaporation would theoretically never conclude. In this case, due to the discrete nature of particle emission, the process would continue for an extraordinarily long time—far exceeding the current age of the universe—until the last particle is emitted.

In the case of a charged Reissner-Nordstr\"om black hole, this scenario does not hold because: (a) a macroscopic black hole cannot sustain a significant electric charge, as it rapidly discharges into the surrounding environment \cite{Gibbons:1975kk}, and (b) once the charge is no longer near-extremal, it radiates away much faster than the mass, ensuring that the final stage of evaporation resembles that of a Schwarzschild black hole \cite{Carter:1974yx,Page:1977um,Novikov:1980ni} (though some scenarios suggest that it may evaporate until the extremely charged state \cite{Hiscock:1990ex}).

At this final stage of evaporation, we encounter the long-standing question: what is the ultimate fate of the black hole? Several alternative hypotheses have been proposed, including the complete evaporation of the black hole with a burst of high-energy radiation at the end, the transformation into a naked singularity, or the existence of a remnant, as suggested by models incorporating the generalized uncertainty principle \cite{Adler:2001vs,Adler:1999bu}. However, all of these scenarios remain speculative because they rely on a complete understanding of the quantum gravity regime.

In this work, we propose a new scenario for black hole evaporation that avoids the need for quantum gravity. We suggest that macroscopic black holes (those much larger than the Planck scale) do not reach the quantum gravity regime as they evaporate, but instead transition into macroscopic wormholes. According to this scenario, black holes would only reach quantum gravity scales if they were initially quantum objects, created either in the early universe or in a laboratory setting.

This conclusion is based on solutions to the field equations that describe different objects depending on the value of a new parameter that governs deviations from the Schwarzschild geometry. Specifically, this parameter describes a regular black hole that transitions into a traversable wormhole through an intermediate extreme state \cite{Simpson:2018tsi,Casadio_2001}. This extreme state is analogous to the extremal Reissner-Nordstr\"om solution, as the Hawking temperature vanishes at the transition point. The new parameter, denoted as $a$, functions similarly to electric charge but is not associated with any physical charge that can radiate away. Instead, it can be interpreted as a coupling parameter of the theory, such as a parameter responsible for tidal forces from extra dimensions.

Since this parameter cannot be radiated away, the ratio $a/M$ increases during evaporation, eventually approaching the extremal limit where $a/M \rightarrow 2$. However, extremal states are typically unstable, and the astrophysical environment, along with the cloud of particles emitted by the black hole, can act as perturbations that push the system out of equilibrium. Thus, we propose that such a near-extremal, very cold black hole cannot remain in equilibrium indefinitely and will eventually transition into a wormhole state.

Furthermore, even if such a near-extremal black hole were stable, it would still remain macroscopic and non-quantum, as the evaporation process would not drive it to quantum scales.

Our paper is organized as follows.
In sec. II we derive wave equations and effective potentials for Maxwell and Dirac fields around general spherically symmetric spacetimes. Sec. III is devoted to essentials of the classical scattering problem and introduction of grey-body factors. Sec. IV gives the basic information on Hawking radiation formula. In sec. V we apply all the above to the two models of black holes: the Simpson-Visser black hole \cite{Simpson:2018tsi} and the Casadio-Fabbri-Mazzacurati (CFM) black hole~\cite{Casadio_2001}. In the end we interpret the obtained results towards the scenario of evaporation of a black hole into a non-quantum wormhole.

\section{The wave-like equations}\label{WE}

It is well known that the emission of matter fields dominates during Hawking radiation, while gravitons contribute approximately one percent or less to the total emission flow \cite{Page:1976df,Page:1976ki}. As a result, considering only the emission of matter fields provides a reliable qualitative picture of Hawking evaporation for various black holes \cite{Konoplya:2023ahd,Konoplya:2020jgt,Konoplya:2020cbv,Konoplya:2019ppy,Konoplya:2019hml}. Therefore, in this work, we restrict our analysis to the Maxwell and Dirac fields, which correspond to the emission of massless elementary particles of the Standard Model and the emission of massive particles in the ultrarelativistic regime, where their mass can be neglected.

For a generic diagonal spherically-symmetrical metric of the form

\begin{align}\label{ABCmetric}
&ds^2=-A(r)dt^2+B(r)dr^2+C(r)d\Omega^2, \\
&\hspace*{5em}d\Omega^2\equiv d\theta^2+\sin^2\theta d\phi^2,\nonumber
\end{align}

The generally covariant equations for the electromagnetic field $A_\mu$ and the massless Dirac field $\Upsilon$ \cite{Brill:1957fx} on the curved background with the metric $g_{\mu\nu}$ are respectively written as:
\begin{subequations}\label{coveqs}
\begin{eqnarray}\label{EmagEq}
\frac{1}{\sqrt{-g}}\partial_{\mu} \left(F_{\rho\sigma}g^{\rho \nu}g^{\sigma \mu}\sqrt{-g}\right)&=&0\,,
\\\label{covdirac}
\gamma^{\alpha} \left( \frac{\partial}{\partial x^{\alpha}} - \Gamma_{\alpha} \right) \Upsilon&=&0,
\end{eqnarray}
\end{subequations}
where $F_{\mu\nu}=\partial_\mu A_\nu-\partial_\nu A_\mu$ is the electromagnetic tensor, $\gamma^{\alpha}$ are gamma matrices in a curved space-time, and $\Gamma_{\alpha}$ are spin connections in the tetrad formalism.
After standard separation of the variables and transition to the ``tortoise coordinate'' $r_*$ equations (\ref{coveqs}) take the Schr\"odinger-like form (see, for instance, \cite{Konoplya:2011qq,Kokkotas:1999bd} and references therein)
\begin{equation}\label{wave-equation}
\dfrac{d^2 \Psi}{dr_*^2}+(\omega^2-V(r))\Psi=0.
\end{equation}
The tortoise coordinate $r_*$ is defined by the relation
\begin{equation}\label{tortoise}
dr_*\equiv\frac{dr}{\sqrt{A(r)/B(r)}}.
\end{equation}
The effective potential for the electromagnetic field is
\begin{equation}\label{Vem}
V_E(r)=\frac{A(r)\ell(\ell+1)}{C(r)},
\end{equation}
where $\ell=1, 2, 3, \ldots$ are the multipole numbers.

\begin{figure}[H]
\centerline{\resizebox{\linewidth}{!}{\includegraphics*{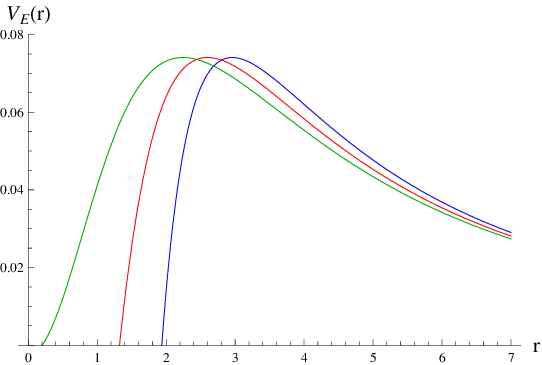}}}
\caption{The effective electromagnetic potential $V_E(r)$ in the Simpson--Visser model for $\ell=1$ and $a/M=0.5, 1.5, 1.99$ (right to left), $M=1$.}
\label{figVe-SV}
\end{figure}

\begin{figure}[H]
\centerline{\resizebox{\linewidth}{!}{\includegraphics*{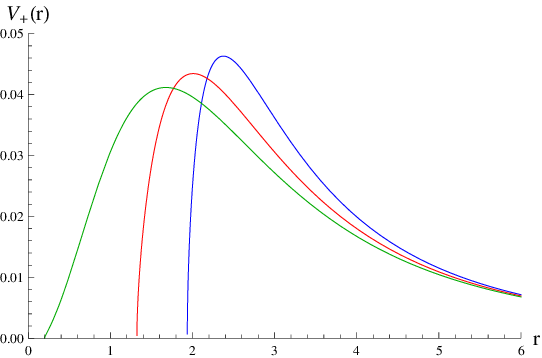}}}
\caption{The effective Dirac potential $V_+(r)$ in the Simpson--Visser model for $\ell=1/2$ and $a/M=0.5, 1.5, 1.99$ (right to left), $M=1$.}
\label{figVdir-SV}
\end{figure}

\begin{figure}[H]
\centerline{\resizebox{\linewidth}{!}{\includegraphics*{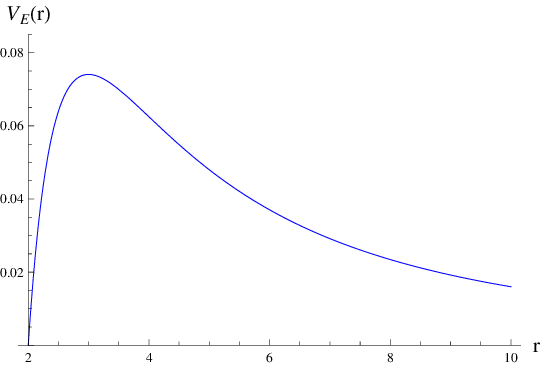}}}
\caption{The effective electromagnetic potential $V_E(r)$ in the CFM model for $\ell=1$, $M=1$. Note that $V_E(r)$ does not depend on values of the parameter $a$.}
\label{figVe-CFM}
\end{figure}

\begin{figure}[H]
\centerline{\resizebox{\linewidth}{!}{\includegraphics*{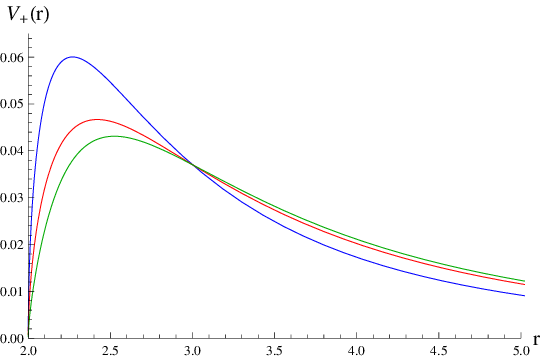}}}
\caption{The effective Dirac potential $V_+(r)$ in the CFM model for $\ell=1/2$ and $a/M=0, 1.5, 1.9$ (from top to bottom), $M=1$.}
\label{figVdir-CFM}
\end{figure}

For the Dirac field  we have two isospectral potentials~\cite{Arbey_2021}
\begin{equation}\label{Vplus}
V_{\pm}(r)=\frac{A(r)k^2}{C(r)}\pm k\frac{d}{dr_*}\sqrt{\frac{A(r)}{C(r)}},
\end{equation}
where the integer $k=1,2,3,...$ is connected with the half-integer fermionic multipole number $l=1/2, 3/2, ...$ as $k=l+1/2$.
%The iso-spectral wave functions can be transformed one into another by the Darboux transformation
%\begin{equation}\label{psi}
%\Psi_{+}=q \left(W+\dfrac{d}{dr_*}\right) \Psi_{-}, \quad q={\rm const}.
%\end{equation}
Due to iso-spectrality, we will use only one of the effective potentials, $V_{+}(r)$, because the WKB method is more accurate for it.
The effective potentials are shown in figs. \ref{figVe-SV}, \ref{figVdir-SV} for the Simpson-Visser model and in figs. \ref{figVe-CFM}, \ref{figVdir-CFM} for the CFM model.

\section{Grey-body factors of black holes}\label{GBfactor}

Frequently, when estimating the intensity of Hawking radiation, grey-body factors are often assumed to be unity, implying that temperature is the dominant factor and that the linear coefficient can be neglected. Nevertheless, the grey-body factor plays a crucial role in determining the fraction of the initial Hawking radiation that is reflected back towards the event horizon by the potential barrier surrounding the black hole. This factor is essential because the radiation emitted by a black hole, as predicted by Hawking, does not directly escape to infinity. Instead, it encounters a potential barrier, which partially reflects the radiation back into the black hole. The grey-body factor quantifies this reflection and transmission process, offering a more precise understanding of the radiation observable by a distant observer. By incorporating the grey-body factor into Hawking's semi-classical formula, one can calculate the actual amount of radiation that will reach infinity. As highlighted in \cite{Konoplya:2019ppy}, the grey-body factor can sometimes have a more significant effect on the radiation spectrum than the black hole's temperature itself, making its calculation vital for accurate predictions.

In this work, we focus on solving the wave equation (\ref{wave-equation}) under boundary conditions that allow for the existence of incoming waves from infinity. This setup is mathematically equivalent to considering waves scattered from the black hole's event horizon due to the symmetry of the scattering process. Therefore, the appropriate boundary conditions for the scattering problem in equation (\ref{wave-equation}) are as follows:
\begin{equation}\label{BC}
\begin{array}{ccll}
    \Psi &=& e^{-i\omega r_*} + R e^{i\omega r_*},& r_* \rightarrow +\infty, \\
    \Psi &=& T e^{-i\omega r_*},& r_* \rightarrow -\infty, \\
\end{array}%
\end{equation}
where $R$ and $T$ represent the reflection and transmission coefficients, respectively. These coefficients encapsulate the behavior of the wave as it interacts with the black hole's surrounding potential barrier.

Analysis of Hawking radiation of various four-dimensional black holes show that gravitons contribute only a small part (less than two percent) into the total flux of radiation \cite{Page:1976df,Page:1976ki,Page:1977um}, so that consideration of matter fields only is sufficient for quite accurate estimation of the intensity of Hawking radiation. 
In addition, emission of massive particles is either occurs ultra-relativistically or greatly suppressed for large masses.
Therefore, here we will be limited by  Maxwell and massless Dirac fields which include Standard Model Particles emitted by a relatively cold black hole, which is much heavier than the Plank mass.

The effective potential for the electromagnetic field takes the form of a potential barrier, which monotonically decreases towards spatial infinity and the event horizon. This structure of the potential allows us to apply the WKB (Wentzel-Kramers-Brillouin) approximation \cite{Schutz:1985km,Iyer:1986np,Konoplya:2003ii}, a semi-classical method used to estimate the reflection and transmission coefficients, $R$ and $T$. Since $\omega^2$ is real, the first-order WKB calculations yield real values for $R$ and $T$ \cite{Schutz:1985km,Iyer:1986np,Konoplya:2003ii}, satisfying the fundamental energy conservation relation:
\begin{equation}\label{1}
\left|T\right|^2 + \left|R\right|^2 = 1.
\end{equation}
This ensures that all the energy from the wave is either transmitted to infinity or reflected back towards the black hole.

For the Dirac field, the situation is slightly more complex due to the non-monotonic behavior of the effective potential near the event horizon for the minus chirality. This non-monotonicity introduces three turning points in the scattering problem, but despite this added complexity, the WKB approximation remains reasonably accurate for this case as well. However, the potentials for plus and minus chiralities are iso-spectral and, thereby, produce the same grey-body factors, so that using the plus-potential in the WKB method is safe.

Once the reflection coefficient $R$ is obtained, the transmission coefficient $T$ for each multipole number $\ell$ can be calculated using the relation:
\begin{equation}
\left|{\pazocal A}_{\ell}\right|^2=1-\left|R_{\ell}\right|^2=\left|T_{\ell}\right|^2.
\end{equation}
This transmission coefficient corresponds to the portion of the wave that successfully escapes to infinity and can be detected by a distant observer.

To achieve a higher degree of accuracy in the calculation of the transmission coefficients, we employ the higher-order WKB approximation (see reviews  \cite{Konoplya:2019hlu,Konoplya:2011qq}). This approach provides a more reliable estimation, particularly for intermediate and high values of $\omega$, where the WKB method performs well. However, for very small $\omega$, which corresponds to nearly complete reflection of the wave, the WKB method becomes less effective. In such cases, the contribution to the total energy emission is minimal, so we can safely use an extrapolation of the WKB results to these smaller frequencies. According to \cite{Schutz:1985km,Iyer:1986np,Konoplya:2003ii}, the reflection coefficient $R$ in the WKB approximation can be expressed as:
\begin{equation}\label{moderate-omega-wkb}
R = (1 + e^{- 2 i \pi K})^{-\frac{1}{2}},
\end{equation}
where $K$ is determined by solving the following equation:
\begin{equation}
K - i \frac{(\omega^2 - V_{max})}{\sqrt{-2 V_{max}^{\prime \prime}}} - \sum_{i=2}^{i=6} \Lambda_{i}(K) =0,
\end{equation}
In this expression, $V_{max}$ is the maximum of the effective potential, $V_{max}^{\prime \prime}$ is its second derivative with respect to the tortoise coordinate, and $\Lambda_i$ are the higher-order WKB correction terms. The WKB series is not guaranteed to converge at every order; instead, it converges asymptotically. Therefore, there exists an optimal order at which the accuracy of the WKB method is the highest. This optimal order depends on the specific shape of the effective potential. In our case, we used the 6th order WKB approximation for studying perturbations of the Maxwell field and the Dirac field with the plus-potential. These choices are motivated by their proven accuracy in the Schwarzschild black hole limit and number of its generalizations/modifications (see for instance recent works \cite{Konoplya:2023ahd,Stashko:2024wuq,Bolokhov:2024ixe,Bolokhov:2023bwm,Malik:2024nhy,Malik:2023bxc,Skvortsova:2023zmj,Skvortsova:2024wly,Konoplya:2024lch,Dubinsky:2024aeu,Yang:2024ofe,Konoplya:2020bxa,Dubinsky:2024hmn} and references therein), and we expect similar behavior for the regular black hole models we are considering.

It is also worth of mentioning that the grey-body factors obtained here can be found with reasonable accuracy from the quasinormal of black holes \cite{Bronnikov:2019sbx,Churilova:2019cyt,Malik:2024qsz,Malik:2024itg} owing to the correspondence between quasinormal modes and grey- body factor established in \cite{Konoplya:2024lir,Konoplya:2024vuj}.

\section{Intensity of Hawking radiation}\label{HawkingR}

In the following analysis, we assume that the black hole remains in a state of thermal equilibrium with its surroundings. This implies that the black hole's temperature remains constant between the emission of successive particles during the Hawking radiation process. Such an assumption simplifies the description of the system, allowing it to be treated within the framework of the canonical ensemble. The use of the canonical ensemble in black hole thermodynamics is well established and has been extensively discussed in the literature (see, for instance, \cite{Kanti:2004nr} for a comprehensive review). 

Under this assumption of thermal equilibrium, the familiar formula for the energy emission rate of Hawking radiation can be applied. This formula, originally derived by Hawking \cite{Hawking:1975vcx}, relates the energy emitted by the black hole per unit time to the temperature, the grey-body factors, and the frequency spectrum of the emitted radiation:
\begin{align}\label{energy-emission-rate}
\frac{\text{d}E}{\text{d} t} = \sum_{\ell}^{} N_{\ell} \left| \pazocal{A}_l \right|^2 \frac{\omega}{\exp\left(\omega/T_\text{H}\right)\pm1} \frac{\text{d} \omega}{2 \pi},
\end{align}
where $T_H$ is the Hawking temperature, $\pazocal{A}_l$ are the grey-body factors (which account for the scattering of radiation in the curved spacetime around the black hole), and $N_\ell$ represents the multiplicities associated with each angular momentum mode $\ell$, which depend on the number of particle species and other properties such as polarizations and helicities. The sum is taken over all angular momentum modes, and the $\pm 1$ term corresponds to the nature of the particles being emitted, with the plus sign for fermions and the minus sign for bosons.

\begin{figure}[H]
\centerline{\resizebox{\linewidth}{!}{\includegraphics*{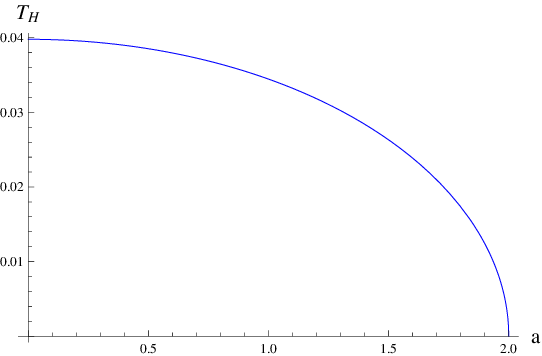}}}
\caption{The Hawking temperature of the Simpson--Visser black hole as a function of $a\in[0, 2M]$, $M=1$.}
\label{figTH-SV}
\end{figure}

\begin{figure}[H]
\centerline{\resizebox{\linewidth}{!}{\includegraphics*{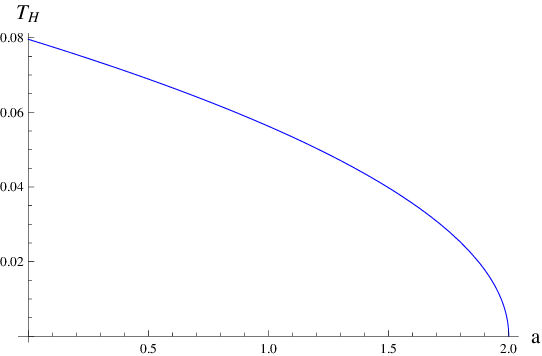}}}
\caption{The Hawking temperature of the CFM black hole as a function of $a\in[0, 2M]$, $M=1$.}
\label{figTH-CFM}
\end{figure}

As mentioned before the grey-body factors, $\pazocal{A}_l$, play a crucial role in modifying the energy spectrum of the radiation as they account for the probability that radiation emitted near the black hole’s event horizon successfully escapes to infinity. Without this factor, one would incorrectly assume that all emitted radiation reaches a distant observer, ignoring the potential barrier that partially reflects the radiation back into the black hole.

In our analysis, we also consider the dependence of the Hawking temperature on quantum deformations of the black hole spacetime, such as the tidal parameter $a$. Figures \ref{figTH-SV} and \ref{figTH-CFM} show how the Hawking temperature decreases as the tidal parameter $a$ increases, indicating that as these deformations grow stronger, the black hole radiates less energy. This behavior is observed for both of the black hole models under consideration.

The multiplicity factors, $N_\ell$, are determined by the number of degenerate $m$-modes for a given angular momentum $\ell$, along with the number of particle species and their respective degrees of freedom (e.g., polarizations and helicities). In the case of four-dimensional spherically symmetric black holes, the degeneracy of the $m$-modes is given by $m = -\ell, -\ell+1, \dots, \ell-1, \ell$, resulting in $2\ell + 1$ modes for each $\ell$. The multiplicity factor also accounts for the number of particle species, which varies depending on the field type. Thus, for Maxwell (electromagnetic) and Dirac (fermionic) fields, the multiplicity factors $N_\ell$ take the following forms:
\begin{align}
&N_{\ell} = 2 (2 \ell+1),\quad  \ell=1, 2, 3, \dots \qquad (\rm Maxwell),\\
&N_{l} = 36k, \qquad \quad \,\, k=1, 2, 3, \dots \qquad (\rm Dirac).
\end{align}
The formulae above can be explicitly derived. For the Maxwell field, we have $N_{\ell}=d_{\rm Maxwell}\times(2\ell+1)$, where $\ell=1,2,3,...$, and $d_{\rm Maxwell}=2$ is the number of degrees of freedom for a photon's state which has only two distinguishable polarizations and is a truly neutral particle identical to its own antiparticle. For the Dirac field, one has $N_l=d_{\rm Dirac}\times(2l+1)$, $l=1/2, 3/2,...$, where a multiplier is derived from the number of degrees of freedom for the leptons in the Standard Model in the ultrarelativistic regime, namely:
$d_{\rm Dirac}$=[3 generations of leptons ($e, \mu, \tau$) $\times$ 2 (antiparticle accounting) $\times$ 2 (spin projections)] + [3 generations of neutrino ($\nu_e, \nu_\mu, \nu_\tau$) $\times$ 2 (antiparticle accounting) $\times$ (1 observable helicity type)]=18. In terms of integer $k=l+1/2=1,2,3...$ we thus have $N_l=d_{Dirac}\times(2l+1)=d_{Dirac}\times 2k=36k$. Additionally, both the "plus" and "minus" potentials associated with the Dirac field are related through Darboux transformations, leading to an isospectral problem where both chiralities exhibit the same grey-body factors. This symmetry simplifies the calculation, as it ensures that the emission from both chiralities is governed by identical transmission probabilities.
Notice that in the seminal works by Don Page \cite{Page:1976df,Page:1976ki} the third generation of leptons was not taken into account. 

As the black hole radiates energy through Hawking radiation, its mass decreases over time. The rate at which the black hole loses mass can be expressed using the following well-known formula derived by Page \cite{Page:1976df}:
\begin{equation}\label{dMdt}
\frac{d M}{d t} = -\frac{\hbar c^4}{G^2} \frac{\alpha_{0}}{M^2},
\end{equation}
where we have restored the physical constants $\hbar$ (the reduced Planck constant), $c$ (the speed of light), and $G$ (the gravitational constant). In this equation, $\alpha_{0}$ represents the energy emission rate, $\frac{dE}{dt}$, evaluated at the initial mass $M_0$ of the black hole. This relation highlights the inverse square dependence on the black hole mass, meaning that smaller black holes radiate energy at a faster rate, accelerating their evaporation as they lose mass. As the black hole evaporates, its temperature increases, leading to a more rapid emission of radiation, creating a runaway process in the final stages of evaporation.

It is important to note that this formula assumes the black hole is sufficiently large for semi-classical approximations to hold. In the final stages of evaporation, when the black hole approaches Planck-scale sizes, quantum gravity effects are expected to become significant, potentially altering this classical picture of Hawking radiation.

\begin{figure}[H]
\centerline{\resizebox{\linewidth}{!}{\includegraphics*{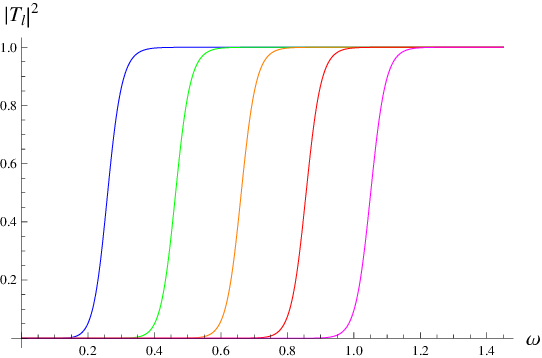}}}
\caption{Grey-body factors of the electromagnetic field per frequency unit for the Simpson--Visser black hole with $a/M=1.99$ for $\ell=1,2,3,4,5$ (left to right).}
\label{figGB-EM-a1_99-SV}
\end{figure}

\begin{figure}[H]
\centerline{\resizebox{\linewidth}{!}{\includegraphics*{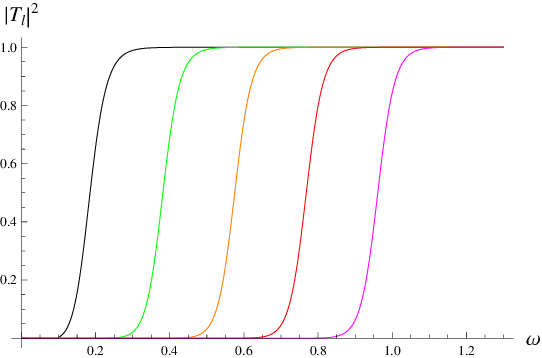}}}
\caption{Grey-body factors of the Dirac field per frequency unit for the Simpson--Visser black hole with $a/M=1.99$ for $\ell=1/2, 3/2, 5/2, 7/2, 9/2$ (left to right).}
\label{figGB-Dir-a1_99-SV}
\end{figure}

\section{Regular black hole - wormhole transition}

\subsection{Simpson-Visser model}

The Simpson--Visser model proposed in \cite{Simpson:2018tsi} is an example of a spherically-symmetrical spacetime whose geometric properties are governed by a certain regularizing parameter $a$. This spacetime interpolates between Schwarzschild black hole (BH) and a traversable wormhole, having a black bounce and an extremal null-bounce as intermediate stages, when increasing the value of the parameter $a$. The Simpson--Visser model corresponds to $A(r)=1/B(r)=f(r)$ and $C(r) =r^2+a^2$, that is the metric has the form
\begin{equation}
ds^2=-f(r)dt^2+\frac{dr^2}{f(r)}+(r^2+a^2)d\Omega^2,
\end{equation}
where 
\begin{equation}
f(r)=1-\frac{2M}{\sqrt{r^2+a^2}}.
\end{equation}
This metric is regular everywhere when $a\neq 0$.

The geometry of Simpson--Visser's spacetime admits regions with $r\in\mathbb{R}$, but for our purposes we restrict ourselves to considering the region $r\ge0$. Without loss of generality one can assume $a\ge0$. The BH horizon exists when $a<2M$, and is situated at $r_0=\sqrt{4M^2-a^2}$. At the critical (``extreme'') value $a=2M$, one has a one-way wormhole with an extremal null throat located at $r=0$, whereas the value $a>2M$ yields a traversable Morris-Thorne wormhole \cite{Simpson:2018tsi}. Further properties of this spacetime and its generalizations have been considered in \cite{Simpson:2019mud,Simpson:2019cer,Lobo:2020ffi,Mazza:2021rgq,Churilova:2019cyt,Nascimento:2020ime}, while the source term producing this metric has been suggested in \cite{Bronnikov:2021uta}.

\begin{figure}[H]
\resizebox{\linewidth}{!}{\includegraphics*{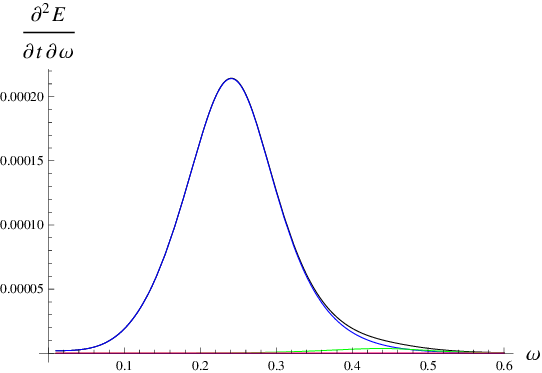}}
\caption{Energy emission rate of the electromagnetic field per frequency unit for the $a=0$ (Schwarzschild limit) in the Simpson--Visser model for various values of $\ell$: black (top, total), blue ($\ell=1$), green ($\ell=2$). Emission rates for higher $\ell$ are strongly suppressed and invisible in the plot.}\label{figRate-EM-a0-SV}
\end{figure}

\begin{figure}[H]
\resizebox{\linewidth}{!}{\includegraphics*{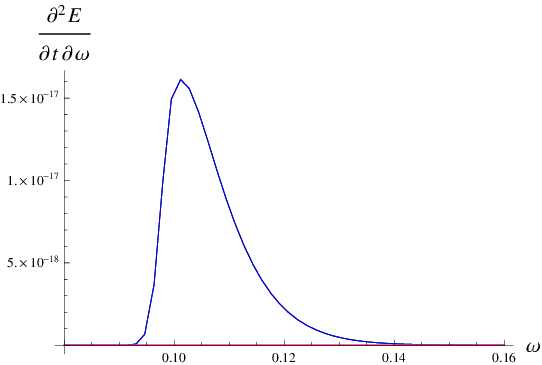}}
\caption{Energy emission rate of the electromagnetic field per frequency unit for the $a/M=1.99$ in the Simpson--Visser model for various $\ell$. Here the blue curve corresponds to $\ell=1$, and the emission rates for higher $\ell$ are strongly suppressed and invisible in the plot.}\label{figRate-EM-a1_99-SV}
\end{figure}

\begin{figure}[H]
\resizebox{\linewidth}{!}{\includegraphics*{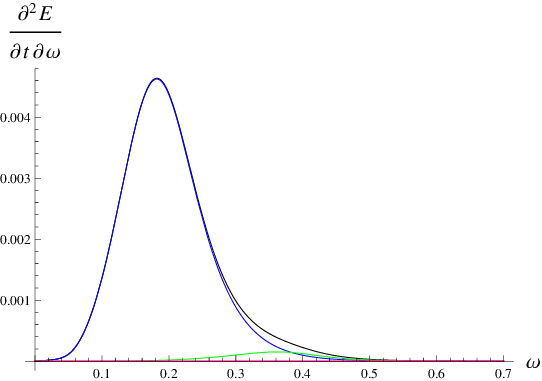}}
\caption{Energy emission rate of the Dirac field per frequency unit for the $a=0$ (Schwarzschild limit) in the Simpson--Visser model for various values of $\ell$: black (top, total), blue ($\ell=1$), green ($\ell=2$). Emission rates for higher $\ell$ are strongly suppressed and invisible in the plot.}\label{figRate-Dir-a0-SV}
\end{figure}

\begin{figure}[H]
\resizebox{\linewidth}{!}{\includegraphics*{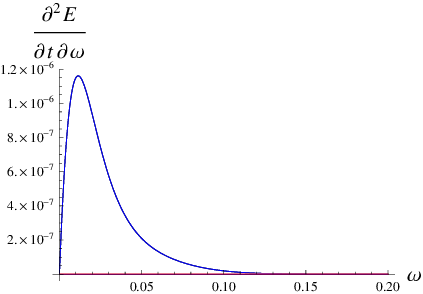}}
\caption{Energy emission rate of the Dirac field per frequency unit for the $a/M=1.96$ in the Simpson--Visser model for various $\ell$. Here the blue curve corresponds to $\ell=1$, and the emission rates for higher $\ell$ are strongly suppressed and invisible in the plot.}\label{figRate-Dir-a1_96-SV}
\end{figure}

The Hawking temperature \cite{Hawking:1975vcx} in units $\hbar=k_{\rm B}=1$ is
\begin{equation}
T_H = \left.\frac{f'(r)}{4 \pi}\right\vert_{r=r_{0}}\!\!=
\frac{\sqrt{4M^2-a^2}}{16\pi M^2}.
\end{equation}

Using Eqs.~\eqref{ABCmetric}--\eqref{Vplus}, one can find the effective potentials for the Simpson--Visser model. The plots of these potentials, as well as the Hawking temperature, are depicted in Figs.~\ref{figVe-SV}, \ref{figVdir-SV}, \ref{figTH-SV}.

Utilizing the technique described in Sec.~\ref{GBfactor} and \ref{HawkingR}, we calculated grey-body factors and energy emission rates for this model at various values of $\ell$ and $a$. The results are shown in Figs. \ref{figGB-EM-a1_99-SV}, \ref{figGB-Dir-a1_99-SV},  \ref{figRate-EM-a0-SV}, \ref{figRate-EM-a1_99-SV}, \ref{figRate-Dir-a0-SV}, \ref{figRate-Dir-a1_96-SV}. Table~\ref{TableSV} represents the total energy emission rate $dE/dt$ for Maxwell and Dirac particles.

During the evaporation of a black hole and decreasing its mass $M$, the ratio $a/M$ will increase up to its extreme value: $a/M\to2$. The supposition is that, despite  decreasing $T_H$ and energy emission rate during this process, the reached near-extremal state should not be expected to be ``frozen'' due to the presence of various microscopic fluctuations, and should eventually lead to the formation of a wormhole. This is more than just a hypothesis, as it is well established that extreme black holes of various types (such as extremally charged or extremally rotating black holes) are unstable \cite{Aretakis:2012ei,Aretakis:2011ha,Lucietti:2012sf,Lucietti:2012xr}. There are also examples of instabilities of near-extreme black holes \cite{Konoplya:2008au}. This suggests that a large, near-extreme black hole in an astrophysical environment should not remain stable in this near-extreme state indefinitely, but, under influence of the environment, rather transition to another compact object or disintegrate. The transition to another compact object would likely result in either a singularity, which is usually also unstable \cite{Gleiser:2006yz,Dotti:2006gc} and expected to decay further, or, more naturally, to a wormhole, which is described here by the same spacetime, depending on one free parameter. The astrophysical environment that could disturb the equilibrium of a near-extreme black hole includes not only the usual matter surrounding black holes, such as an accretion disk, but also the cloud of particles emitted by the black hole during the evaporation process.    

Integrating Eq.~\eqref{dMdt}, one can obtain the decay time of a BH in a certain mass interval $[M_0, M_1]$ taking into account that the coefficient $\alpha$ generally depends on a value of the mass:
\begin{equation}\label{tau}
\tau=-\frac{G^2}{\hbar c^4}\int\limits_{M_0}^{M_1}\frac{M^2dM}{\alpha(M)}=
\frac{G^2a^3}{\hbar c^4}\int\limits_{\xi_0}^{\xi_1}\frac{d\xi}{\xi^4\alpha(\xi)}, \quad \xi\equiv a/M.
\end{equation}
In the last expression, we express the time in terms of the growing ratio $\xi\equiv a/M$ assuming $a=\rm const$. The coefficient $G^2/(\hbar c^4)\simeq5.22976\times10^{-30} {\,\rm s\cdot g^{-3}}$, and a BH mass is supposed to be expressed in grams.

Consider an example when a BH starts decaying from a state with $\xi_0=a/M_0=1$ (that is, $a$ is assumed to be equal to an initial BH mass $M_0$) to the near-extreme state $\xi_1\to2$ thereby reducing its mass by half. For estimation purposes, one can use a rough discrete approximation of the ``half-decay'' time $\tau_{1/2}$ as a sum over set of $\xi_i$ and $\alpha(x_i)$ values according to the numerical data in Table~\ref{TableSV} (where the coefficient $\alpha$ is the sum of the Maxwell and Dirac total energy emission rate contributions):
\begin{equation}\label{tau-discr}
\tau_{1/2}=\frac{G^2M_0^3}{\hbar c^4}\int\limits_{1}^{\xi_1\to2}\!\!\frac{d\xi}{\xi^4\alpha(\xi)}\sim
\frac{G^2M_0^3}{\hbar c^4}\sum\limits_{i}\frac{\Delta\xi_i}{\xi_i^4\alpha(\xi_i)}.
\end{equation}
Due to rapid decreasing of the emission rate coefficient  $\alpha(\xi)$ at $\xi\to2$, only last terms in the sum mostly contribute. 

For the Simpson-Visser model, one can thus obtain an approximate ``half-decay'' time for various values of the initial BH mass $M_0$. The results are shown in Table~\ref{TableSV-tau}.

\begin{table}[H]
\centering
\caption{
	The total energy emission rate $dE/dt$ for Maxwell and Dirac particles 
	in the Simpson--Visser model for various values of the parameter $a$. Summing over $\ell$ (in the Maxwell case) or $k$ (in the Dirac case) is done for the first five multipole values. For brevity, only first three of them are depicted. The power is in units 
     $\hbar c^{6} G^{-2} M^{-2}= 1.719 \cdot 10^{50} (M/\rm g)^{-2}\,\rm erg \cdot s^{-1}$.
     %The last column indicates an estimated lifetime of the black hole with mass $M\sim 10^{15} {\rm g}$.
     }
\label{TableSV}
\medskip     
\begin{tabular}{|l|l|l|l|}
  \hline 
  \hline 
\text{a/M} & \text{$\ell $ or k} & \text{Maxwell} & \text{Dirac} %& \text{$\tau $, sec} 
\\  \hline 
%--- WKB6 Maxwell, WKB6 Dirac ---
\text{      0} & \text{      1} & \text{ 0.0000335107} & \text{ 0.000668252} \\
 \text{      0} & \text{      2} & \text{ 6.67916}$\times 10^{-7}$ & \text{ 0.0000270603} \\
 \text{      0} & \text{      3} & \text{ 1.00693}$\times 10^{-8}$ & \text{ 5.32304}$\times 10^{-7}$ \\
 \text{      0} & \text{(total)} & \text{ 0.0000341888} & \text{ 0.000695853} \\ \hline 
 \text{     0.5} & \text{      1} & \text{ 0.0000267836} & \text{ 0.000564719} \\
 \text{     0.5} & \text{      2} & \text{ 4.57715}$\times 10^{-7}$ & \text{ 0.0000198581} \\
 \text{     0.5} & \text{      3} & \text{ 5.90719}$\times 10^{-9}$ & \text{ 3.35485}$\times 10^{-7}$ \\
 \text{     0.5} & \text{(total)} & \text{ 0.0000272473} & \text{ 0.000584917} \\ \hline 
 \text{      1} & \text{      1} & \text{ 0.0000118927} & \text{ 0.00031027} \\
 \text{      1} & \text{      2} & \text{ 1.15213}$\times 10^{-7}$ & \text{ 6.42987}$\times 10^{-6}$ \\
 \text{      1} & \text{      3} & \text{ 8.37968}$\times 10^{-10}$ & \text{  6.1987}$\times 10^{-8}$ \\
 \text{      1} & \text{(total)} & \text{ 0.0000120087} & \text{ 0.000316763} \\ \hline 
 \text{     1.5} & \text{      1} & \text{ 1.29303}$\times 10^{-6}$ & \text{ 0.0000625938} \\
 \text{     1.5} & \text{      2} & \text{ 2.54734}$\times 10^{-9}$ & \text{ 2.90341}$\times 10^{-7}$ \\
 \text{     1.5} & \text{      3} & \text{ 3.68553}$\times 10^{-12}$ & \text{ 5.76425}$\times 10^{-10}$ \\
 \text{     1.5} & \text{(total)} & \text{ 1.29558}$\times 10^{-6}$ & \text{ 0.0000628847} \\ \hline 
 \text{     1.9} & \text{      1} & \text{ 1.70161}$\times 10^{-9}$ & \text{ 3.27832}$\times 10^{-7}$ \\
 \text{     1.9} & \text{      2} & \text{ 7.17893}$\times 10^{-15}$ & \text{ 1.73694}$\times 10^{-11}$ \\
 \text{     1.9} & \text{      3} & \text{ 1.83967}$\times 10^{-19}$ & \text{ 2.01512}$\times 10^{-16}$ \\
 \text{     1.9} & \text{(total)} & \text{ 1.70162}$\times 10^{-9}$ & \text{  3.2785}$\times 10^{-7}$ \\ \hline 
 \text{    1.99} & \text{      1} & \text{ 2.35335}$\times 10^{-19}$ & \text{ 5.86923}$\times 10^{-9}$ \\
 \text{    1.99} & \text{      2} & \text{   1.478}$\times 10^{-25}$ & \text{ 5.73446}$\times 10^{-14}$ \\
 \text{    1.99} & \text{      3} & \text{ 2.31324}$\times 10^{-32}$ & \text{ 1.82544}$\times 10^{-26}$ \\
 \text{    1.99} & \text{(total)} & \text{ 2.35335}$\times 10^{-19}$ & \text{ 5.86929}$\times 10^{-9}$
\\ \hline 
\hline 
\end{tabular}
\end{table} 

\begin{table}[H]
\centering
\caption{The approximate ``half-decay'' time $\tau_{1/2}$ of the Simpson-Visser BH transition to the near-extreme state $a/M\sim1.99$ starting with an initial mass $M_0$ and $a/M_0=1$.}
\label{TableSV-tau}
\medskip     
\begin{tabular}{|c|c|}
  \hline 
  \hline 
\text{Initial BH mass $M_0$, g} & \text{ $\tau_{1/2}$, sec} \\  \hline 
 $10^0$ & \text{ 5.60873}$\times 10^{-24}$ \\
 $10^5$ & \text{ 5.60873}$\times 10^{-9}$ \\
 $10^{10}$ & \text{ 5.60873}$\times 10^6$ \\
 $10^{15}$ & \text{ 5.60873}$\times 10^{21}$\\
 \hline 
 \hline 
\end{tabular}
\end{table}

\begin{figure}[H]
\centerline{\resizebox{\linewidth}{!}{\includegraphics*{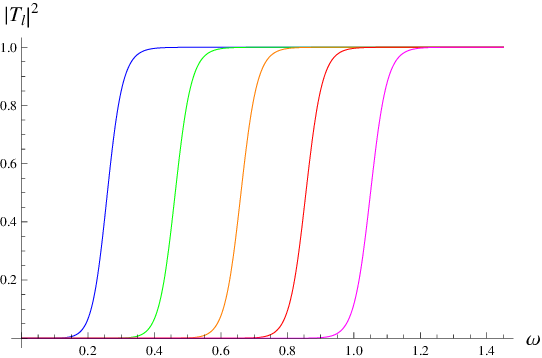}}}
\caption{Grey-body factors of the electromagnetic field per frequency unit for the CFM black hole with $a/M=1.99$ for $\ell=1,2,3,4,5$ (left to right).}
\label{figGB-EM-a1_99-CFM}
\end{figure}

\begin{figure}[H]
\centerline{\resizebox{\linewidth}{!}{\includegraphics*{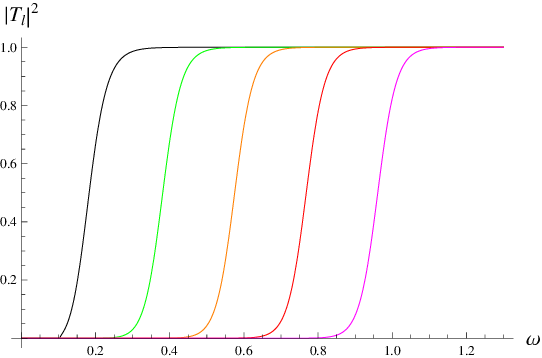}}}
\caption{Grey-body factors of the Dirac field per frequency unit for the CFM black hole with $a/M=1.99$ for $\ell=1/2, 3/2, 5/2, 7/2, 9/2$ (left to right).}
\label{figGB-Dir-a1_99-CFM}
\end{figure}

\subsection{Brane-world models}

Consider one more example based on the so-called Casadio-Fabbri-Mazzacurati (CFM) metric~\cite{Casadio_2001}, which represents a family of black hole / wormhole solutions in the brane-world scenarios and can be used for a possible description of geometry outside a homogeneous star on a brane~\cite{Germani_2001}.

The CFM metric has the following form:
\begin{equation}\label{CFM}
ds^2=-\left(1-\frac{2M}{r}\right)dt^2+\frac{[1-3M/(2r)]dr^2}{(1-2M/r)(1-a/r)}+r^2d\Omega^2.
\end{equation}
This is a family of solutions governed by a parameter $a>0$.  In the case $a<2M$ the metric describes a black hole with a single horizon at $r_0=2M$. In particular, $a=3M/2$ corresponds to the Schwarzschild BH. In the critical (``extreme'') case $a=2M$ there is a black hole with a double horizon at $r=2M$, whereas the values $a>2M$ yield a symmetric traversable wormhole. Quasinormal modes of these black holes have been studied in \cite{Abdalla:2006qj,Malik:2024itg,Bronnikov:2019sbx}.

\begin{figure}[H]
\resizebox{\linewidth}{!}{\includegraphics*{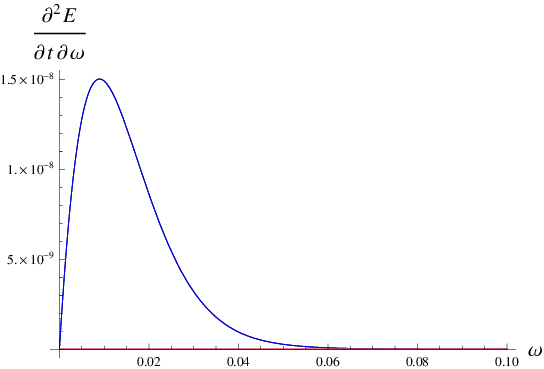}}
\caption{Energy emission rate of the electromagnetic field per frequency unit for the $a/M=1.99$ in the CFM model for various $\ell$. Here the blue curve corresponds to $\ell=1$, and the emission rates for higher $\ell$ are strongly suppressed and invisible in the plot.}\label{figRate-EM-a1_99-CFM}
\end{figure}

\begin{figure}[H]
\resizebox{\linewidth}{!}{\includegraphics*{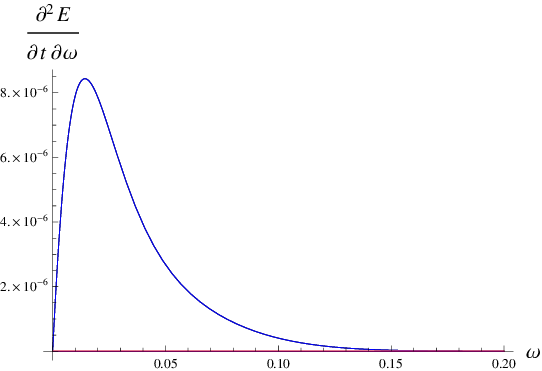}}
\caption{Energy emission rate of the Dirac field per frequency unit for the $a/M=1.97$ in the CFM model for various $\ell$. Here the blue curve corresponds to $\ell=1$, and the emission rates for higher $\ell$ are strongly suppressed and invisible in the plot.}\label{figRate-Dir-a1_97-CFM}
\end{figure}

The Hawking temperature of the CFM black hole in units $\hbar=k_{\rm B}=1$ is
\begin{equation}
T_H = \left.-\frac{1}{4\pi\sqrt{-g_{tt}g_{rr}}}\frac{dg_{tt}}{dr}\right\vert_{r=r_0}=\frac{\sqrt{2-a/M}}{4\sqrt{2}\pi M}.
\end{equation}
The plot of $T_H(a)$ for $M=1$ is depicted in Fig.~\ref{figTH-CFM}.

Figs.~\ref{figVe-CFM} and \ref{figVdir-CFM} show the effective potentials $V_E(r)$ and $V_{+}(r)$ for the Maxwell and Dirac fields calculated on the CFM metric background using Eqs.~\eqref{ABCmetric}--\eqref{Vplus}.

As in the previous section, we have calculated grey-body factors and energy emission rates for electromagnetic and Dirac fields on the CFM BH background at various values of $\ell$ and $a$. The results are represented in Figs. \ref{figGB-EM-a1_99-CFM}, \ref{figGB-Dir-a1_99-CFM},   \ref{figRate-EM-a1_99-CFM}, \ref{figRate-Dir-a1_97-CFM}. Table~\ref{TableCFM} represents the total energy emission rate $dE/dt$ in the Maxwell and Dirac cases.

\begin{table}[H]
\centering
\caption{
	The total energy emission rate $dE/dt$ for Maxwell and Dirac particles 
	in the CFM model for various values of the parameter $a$. Summing over $\ell$ (in the Maxwell case) or $k$ (in the Dirac case) is done for the first five multipole values. For brevity, only first three of them are depicted. The power is in units 
     $\hbar c^{6} G^{-2} M^{-2}= 1.719 \cdot 10^{50} (M/\rm g)^{-2}\,\rm erg \cdot s^{-1}$.
     %The last column indicates an estimated lifetime of the black hole with mass $M\sim 10^{15} {\rm g}$.
     }
\label{TableCFM}
\medskip     
\begin{tabular}{|l|l|l|l|}
%--- WKB6 Maxwell, WKB6 Dirac ---
  \hline 
  \hline 
\text{a/M} & \text{$\ell $ or k} & \text{Maxwell} & \text{Dirac} %& \text{$\tau $, sec} 
\\  \hline 
\text{      0} & \text{      1} & \text{ 0.00227645}
   & \text{ 0.00646277} \\
 \text{      0} & \text{      2} & \text{ 0.00031455}
   & \text{ 0.00392257} \\
 \text{      0} & \text{      3} & \text{
   0.0000504834} & \text{ 0.000886073} \\
 \text{      0} & \text{(total)} & \text{ 0.00264958}
   & \text{ 0.0114327} \\ \hline
 \text{      1} & \text{      1} & \text{ 0.000292203}
   & \text{ 0.00212388} \\
 \text{      1} & \text{      2} & \text{
   0.0000215693} & \text{ 0.000472163} \\
 \text{      1} & \text{      3} & \text{
   1.30613}$\times 10^{-6}$ & \text{ 0.0000364152} \\
 \text{      1} & \text{(total)} & \text{ 0.00031515}
   & \text{ 0.00263475} \\ \hline
 \text{     1.5} & \text{      1} & \text{
   0.0000335107} & \text{ 0.000668252} \\
 \text{     1.5} & \text{      2} & \text{
   6.67916}$\times 10^{-7}$ & \text{ 0.0000270603} \\
 \text{     1.5} & \text{      3} & \text{
   1.00693}$\times 10^{-8}$ & \text{ 5.32304} $\times 10^{-7}$ \\
 \text{     1.5} & \text{(total)} & \text{
   0.0000341888} & \text{ 0.000695853} \\ \hline
 \text{     1.8} & \text{      1} & \text{
   1.00788}$\times 10^{-6}$ & \text{ 0.000058485} \\
 \text{     1.8} & \text{      2} & \text{
   1.52189}$\times 10^{-9}$ & \text{ 1.93097} $\times10^{-7}$ \\
 \text{     1.8} & \text{      3} & \text{ 
   1.7238}$\times 10^{-12}$ & \text{ 3.00646} $\times 10^{-10}$ \\
 \text{     1.8} & \text{(total)} & \text{ 
   1.0094}$\times 10^{-6}$ & \text{ 0.0000586784} \\ \hline
 \text{     1.9} & \text{      1} & \text{
   6.07248}$\times 10^{-8}$ & \text{ 6.98115}$\times 10^{-6}$ \\
 \text{     1.9} & \text{      2} & \text{
   8.21348}$\times 10^{-12}$ & \text{ 2.89974} $\times 10^{-9}$ \\
 \text{     1.9} & \text{      3} & \text{
   1.07615}$\times 10^{-15}$ & \text{ 5.15406} $\times 10^{-13}$ \\
 \text{     1.9} & \text{(total)} & \text{
   6.07331}$\times 10^{-8}$ & \text{ 6.98405}$\times 10^{-6}$ \\ \hline
 \text{    1.95} & \text{      1} & \text{ 
   9.0106}$\times 10^{-9}$ & \text{ 1.02027}$\times 10^{-6}$ \\
 \text{    1.95} & \text{      2} & \text{
   3.25544}$\times 10^{-13}$ & \text{ 1.48107}$\times 10^{-10}$ \\
 \text{    1.95} & \text{      3} & \text{
   2.03765}$\times 10^{-17}$ & \text{ 9.31059}$\times 10^{-15}$ \\
 \text{    1.95} & \text{(total)} & \text{
   9.01092}$\times 10^{-9}$ & \text{ 1.02042}$\times 10^{-6}$ \\ \hline
 \text{    1.97} & \text{      1} & \text{
   3.77518}$\times 10^{-9}$ & \text{  3.5046}$\times  10^{-7}$ \\
 \text{    1.97} & \text{      2} & \text{  
   1.125}$\times 10^{-13}$ & \text{ 4.55415}$\times 10^{-11}$ \\
 \text{    1.97} & \text{      3} & \text{ 
   7.1582}$\times 10^{-18}$ & \text{ 2.80572}$\times 10^{-15}$ \\
 \text{    1.97} & \text{(total)} & \text{
   3.77529}$\times 10^{-9}$ & \text{ 3.50506}$\times 10^{-7}$ \\ \hline
 \text{    1.99} & \text{      1} & \text{
   9.17785}$\times 10^{-10}$ & \text{ 7.16286}$\times 10^{-8}$ \\
 \text{    1.99} & \text{      2} & \text{
   2.14017}$\times 10^{-14}$ & \text{ 8.50583}$\times 10^{-12}$ \\
 \text{    1.99} & \text{      3} & \text{
   1.17875}$\times 10^{-18}$ & \text{ 4.61572}$\times 10^{-16}$ \\
 \text{    1.99} & \text{(total)} & \text{
   9.17806}$\times 10^{-10}$ & \text{ 7.16371}$\times 10^{-8}$ \\
\hline 
\hline 
\end{tabular}
\end{table} 

To estimate an approximate time for reaching such a near-extreme transition regime, we consider again the process of the BH decaying from a state with $a=M_0$ (where $M_0$ is an initial BH mass) to the near-extreme state, reducing the BH mass by half. Applying the approach based on Eqs.~\eqref{tau} and \eqref{tau-discr} and using the numerical data from Table~\ref{TableCFM}, we obtain the corresponding ``half-decay'' time $\tau_{1/2}$ in the CFM model, whose values are shown in Table~\ref{TableCFM-tau} for various initial BH masses.

\begin{table}[H]
\centering
\caption{
	The approximate ``half-decay'' time $\tau_{1/2}$ of the CFM black hole transition to the near-extreme state $a/M\sim1.99$ starting with an initial mass $M_0$ and $a/M_0=1$.
     }
\label{TableCFM-tau}
\medskip     
\begin{tabular}{|c|c|}
  \hline 
  \hline
\text{Initial BH mass $M_0$, g} & \text{ $\tau_{1/2}$, sec} \\  \hline 
$10^0$ & \text{ 1.38002}$\times 10^{-25}$ \\
$10^5$ & \text{ 1.38002}$\times 10^{-10}$ \\
$10^{10}$ & \text{ 138002} \\
$10^{15}$ & \text{ 1.38002}$\times 10^{20}$ \\
 \hline 
 \hline
%\end{array}
\end{tabular}
\end{table}

\section{Discussions}

The final stages of black hole evaporation are closely tied to the long-standing challenge of constructing a consistent theory of quantum gravity. If, due to some modification of Einstein's gravity—such as extra-dimensional scenarios \cite{Casadio_2001} or the addition of matter fields \cite{Simpson:2018tsi} — the black hole acquires an extra parameter, $a$, allowing for an extremal zero-temperature state at a specific ratio of $a/M$, this ratio is expected to increase as the black hole evaporates.

If the initial value of $a/M$ is very small, perhaps tens of orders of magnitude smaller than unity, then for most of its evaporation, the black hole behaves similarly to the classical case described by Einstein’s gravity, eventually approaching the Planck scale while remaining far from the extremal state. On the other hand, if we assume an initial value of $a/M \sim 1$, which is consistent with post-Newtonian constraints, the black hole would need to lose only half of its mass to reach a near-extremal state, where its temperature vanishes and evaporation effectively halts. In this scenario, if the black hole initially had a mass much larger than the Planck mass—i.e., it was still governed by classical (non-quantum) gravity—it would remain non-quantum even at the final stage of evaporation.

Under this assumption, two possibilities arise: (a) primordial black holes with an initial mass greater than the Planck scale could freeze in a near-extremal state and remain non-quantum for a time period far exceeding the age of the universe, or (b) due to the dynamical instability of the extremal state \cite{Aretakis:2012ei,Aretakis:2011ha,Lucietti:2012sf,Lucietti:2012xr}, the black hole could transition into a wormhole with the mass of the same order. Both of these outcomes avoid the complications associated with quantum gravity, the formation of a final singularity, or the emergence of a remnant constrained by the uncertainty principle. The exact initial conditions required for such a transition, however, would need to be investigated in a separate study.

If the initial value of $a/M$ is already close to the extremal limit, then a primordial black hole would have been created cold, and its evaporation would take place in a near-extremal regime, facing a choice between a slow, never-ending evaporation or transitioning into a wormhole state.

It is important to note that the described effect of avoiding the quantum gravity regime during the evaporation process is not limited to the two models considered here. For instance, extremal Hayward and Dymnikova black holes also exhibit vanishing temperature. If these black holes are classified as non-quantum, they would likely follow a similar evaporation scenario.

Our conclusions should also hold when accounting for the effects of backreaction. If the ratio $a/M$ is relatively small at the start of the evaporation process, when the black hole is still large and evaporation occurs in a Schwarzschild-like manner, over time the temperature decreases, and the suppressed radiation results in an even smaller backreaction effect for an almost "frozen" black hole mass.

Even if one does not accept the hypothesis of a transition to a wormhole spacetime, but instead assumes the existence of a modified Schwarzschild solution in some alternative theory of gravity, where a coupling parameter reaches an extremal limit with zero temperature, the argument for the existence of non-quantum, quasi-extremal black holes remains valid. Black holes which were born in the early Universe with the Plank mass would now, most probably, be destroyed in the latest stages of evaporation owing to the dynamical instability of the extreme black holes.

\begin{acknowledgments}
This work was supported by Project No. FSSF-2023-0003.
\end{acknowledgments}

\bibliographystyle{apsrev4-1}
\bibliography{Bibliography}

\end{document}